# Genetically Synthesized Supergain Broadband Wire-Bundle Antenna

Gilad Uziel, Dmytro Vovchuk, Andrey Machnev, Vjačeslavs Bobrovs, and Pavel Ginzburg

*Abstract*— High-gain antennas are essential hardware devices, powering numerous daily applications, including distant point-to-point communications, safety radars, and many others. While a common approach to elevate gain is to enlarge an antenna aperture, highly resonant subwavelength structures can potentially grant high gain performances. The Chu-Harrington limit is a standard criterion to assess electrically small structures and those surpassing it are called superdirective. Supergain is obtained in a case when internal losses are mitigated, and an antenna is matched to radiation, though typically in a very narrow frequency band. Here we develop a concept of a spectrally overlapping resonant cascading, where tailored multipole hierarchy grants both high gain and sufficient operational bandwidth. Our architecture is based on a near-field coupled wire bundle. Genetic optimization, constraining both gain and bandwidth, is applied on a 24-dimensional space and predicts 8.81 dBi realized gain within a half-wavelength in a cube volume. The experimental gain is 6.15 with 13% fractional bandwidth. Small wire bundle structures are rather attractive for designing superscattering and superdirective structures, as they have a sufficient number of degrees of freedom to perform an optimization, and, at the same time rely on simple fabrication-tolerant layouts, based on low-loss materials. The developed approach can be applied to low-frequency (e.g., kHz-MHz) applications, where miniaturization of wireless devices is highly demanded.

## I. INTRODUCTION

ANTENNA ELEMENTS are essential hardware to enable wireless communication links. Among many different characteristics of those devices, gain plays an important role as it has a primary impact on link budget calculations, used to estimate maximally available distances to maintain a link [1]. There are quite a few antenna architectures, developed to achieve high gain. A specific choice of a layout depends on frequency range and other engineering application-specific parameters, constraining the design. In a broader sense, an antenna operation principle can be separated into a pair of categories, i.e., resonant and nonresonant. The latter class, being preferable in terms of bandwidth, scalability, integrability within circuitry, material compatibility, and several other parameters, trades gain properties for a physical aperture size. A typical example here is parabolic reflectors or horns, which gain is directly proportional to the aperture area (A), normalized to an operational wavelength squared ($\lambda 2$) [1]. Resonant antennas, on the other hand, can have a rather small physical aperture (e.g., dipole) and, furthermore, can encompass many resonant elements within a design. Typically, antenna size is assessed versus a radius of a virtual sphere (R), encompassing the structure. Theoretically speaking, electrically small structures with R smaller than $\lambda$ (also, $2R/\lambda<1$ might be used as a criterion) can possess high gain properties [2]. In this case, multiple resonances within a structure should interfere constructively with balanced amplitudes and phases. It is worth noting, that radiation patterns are better described by spherical harmonics (multipoles), forming a complete set of functions on a sphere [3]. However, in structures lacking rotational symmetry, internal resonances are not 1:1 mapped on the multipole expansion, forming the radiation far-field [4]. Nevertheless, it is quite clear that multiple higher-order resonances should be employed in superdirective antenna design. As a result, the structure becomes extremely susceptible to fabrication imperfections, the surrounding environment, and ohmic losses within constitutive materials – all those owing to a significant resonant near-field accumulation within the device. While those aspects are very well known [2], the challenge is to find architectures that are (i) subject to fast optimization, (ii) experimentally realizable, (iii) based on low-loss materials, (iv) can be matched to radiation without employing lossy lumped elements, and (v) provide sufficient operational bandwidth.

Several superdirective antennas were demonstrated, including architectures, based on ceramic resonators (experimentally) [5], [6], multilayer designs (theory) [7]–[9], Huygens sources [10]–[13], (also tuned with lumped elements [14], [15]), and high-impedance surface antennas [16]. Small antenna arrays, with basic elements including, patches, PIFA, monopoles, and others are used for implementing superdirective devices. However, in many cases, an infinite-size (in theory) ground plane serves as an integral part of the system, questioning the validity of applying an electrically small criterion. A promising platform, in application to superdirectivity and superscattering, is small arrays of near-

(Corresponding author: Dmytro Vovchuk e-mail: dimavovchuk@gmail.com).

Gilad Uziel, Dmytro Vovchuk, Andrey Machnev, Pavel Ginzburg School of Electrical Engineering, Tel Aviv University, Ramat Aviv, Tel Aviv, 69978, Israel (e-mail: dimavovchuk@gmail.com, pginzburg@tauex.tau.ac.il ).

Dmytro Vovchuk Department of Radio Engineering and Information Security, Yuriy Fedkovych Fernivtsi National University, Chernivtsi, 58012, Ukraine (e-mail: dimavovchuk@gmail.com)

Vjačeslavs Bobrovs Institute of Telecommunications, Riga Technical University, Riga, LV-1048, Latvia

Gilad Uziel and Dmytro Vovchuk contributed equally to this work.

field coupled resonators [17]–[20], optimized with an aid of fast genetic algorithms.

Here we explore a wire bundle architecture to optimize an antenna resonator for achieving broadband supergain performance. This architecture complies with the five beforehand mentioned requirements. (i) Optimization - a small array of vertically aligned wires is taken as a starting point and then optimized to maximize antenna gain and bandwidth while keeping its footprint small. Each element in the array is allowed to move and have a different length. Those degrees of freedom form a search space. (ii, iii) The realization is based on copper wires, pinched into a Styrofoam. Near-fields are accumulated in a free space between the resonators, making the structure less susceptible to losses in practice. (iv, v) Resonant cascading of interfering multipoles is designed to provide a matching and high directivity over a broad frequency range.

The manuscript is organized as follows: genetic optimization of the structure is presented first and then followed by a detailed electromagnetic analysis to reveal its operation principles. Experimental realization and antenna characterization, along with a comparison to several well-established limits, come before the Conclusion.

## II. Antenna Design

To obtain resonant multipoles overlapping in a subwavelength structure, simultaneous optimization over many degrees of freedom is required. A direct search, in this case, will result in an exponential growth of computational resources, motivating to apply different optimization strategies. Evolutionary algorithms, with genetic optimization being a subset, is a possible compromise, which is intensively explored to design various functional structures. The concept is to treat an electromagnetic configuration as a basic provision in the theory of biological evolution, where processes of selection, mutation, and reproduction govern future development. Evolutionary algorithms [20]–[23] are widely used in multi-dimensional domains, where the functional dependence between parameters is either non-differentiable or has many local extrema. These algorithmic approaches were introduced into engineering problems in the 1960-70s [24] and since then, being supported by ever-growing computational power, started to shift aside from conventional design rules, e.g., [25], [26]. In the context of this report, it is worth mentioning circular rods superscatterers [27], superabsorptive nanoparticles [28], core-shell cylindrical superscatterers [29], subwavelength superscattering nanospheres [30], antenna design, nanoplasmonic particles [31], [32], and others [33]–[39].

Prior to performing an optimization, we will set up our search space. The structure encompasses 9 vertically aligned metal wires. One of them is an active radiating element, which is set fixed. As a standalone, it performs as a dipole antenna, matched to radiate at 6GHz. The rest 8 elements, being initially positioned at the nodes of a symmetric 3x3 array, are allowed to change their position and length independently of each other. This forms an 8x3=24-dimensional search space.

In this realization, the wires are kept mutually parallel. The algorithm flow and the schematic layout of the structure appear in Fig. 1. In brief, the optimization is based on the principles of Genetics and Natural Selection. A population of possible solutions repeatedly undergoes recombination and mutation, producing new children. Each realization is assigned versus a pre-defined fitness function (antenna gain, in our case, while bandwidth was tuned manually), and the best (at this stage) individuals are given a higher chance to evolve. This process keeps repeating until reaching a stopping criterion. Here, an upper limit of 1000 iterations was chosen. The main parts of the algorithm, summarized in the chart, include (i) Selection - selecting individuals, called parents, that contribute to the population of the next generation. The selection is generally stochastic and can depend on the individuals' scores (gain properties here), (ii) Crossover – combining two parents to form children for the next generation, (iii) Mutation – applying random changes to individual parents to form children.

The layout of the optimized structure, which was chosen for the subsequent realization, appears in Table 1. Rows correspond to the wire length and two coordinates within a polar system, linked to the center of the initial array (see Fig. 1(c)). For convenience, the central element is also parametrized, using this approach. Nevertheless, its 2 degrees of freedom (r, φ) are not really independent, as the entire geometry can be linearly shifted and rotated to keep it at the center. The numbering of elements is indicated in Fig. 1(c), where m stays for the center (middle) element and 0 is the feed.

## III. Numerical Analysis and Experimental Verification

To assess the performance of the optimized model, detailed numerical and experimental characterizations will be made next. CST Microwave Studio was used to perform the antenna analysis. The experimental realization of the device is based on copper cut wires, placed in a Styrofoam holder. Fig. 2 (a, b, c) presents photographs from different perspectives. To re-emphasize, the choice of the material platform here is quite important, as superdirective structures are susceptible to material losses and fabrication tolerances. This configuration allows for mitigating both of those constraints. The achieved fabrication accuracies are ±0.03 mm of the wire length and ±0.01 mm in wire positioning, which were obtained by printing the array template on a transparent slide and processing the layout under a table-top microscope. The feeding antenna element is a split wire, soldered to an SMA connector. Balun was not implemented, as its possible contribution here is rather minor, as it will be evident hereinafter. Fig. 2(d) shows the S11- parameter as a function of frequency. A rather good impedance matching (-25 dB around 6 GHz) is observed and predicted numerically. The rest of the panels in Fig. 2 demonstrate the numerically obtained 3D radiation pattern (Fig. 2(e)) and azimuthal and elevation cuts for both numerical and experimental data (panels f and g). Highly-directive shapes are observed.

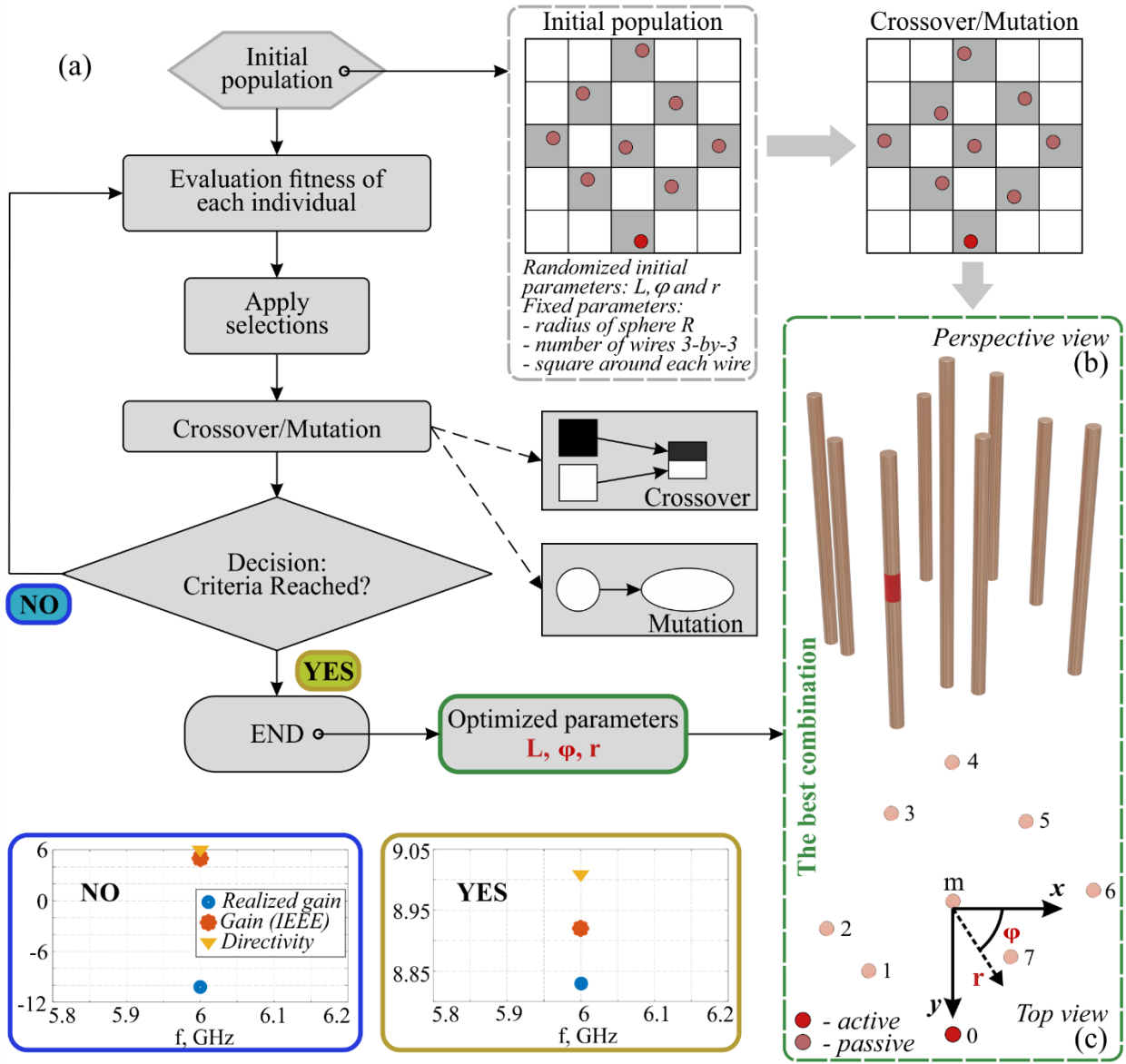

Fig. 1. (a) A flow chart of the superdirective antenna optimization with a genetic algorithm. (b), (c) the structure under optimization – perspective and top view, respectively. Optimization degrees of freedom are indicated in (c). The grey squares show the bounds, restricting the wires' possible positions. A red inset in a wire is a feeding port.

TABLE 1.
The parameters of the optimized antenna, the numbering corresponds to Fig. 1(c).

|   | Center (m) | Feed (0) | 1 | 2 | 3 | 4 | 5 | 6 | 7 |
|---|---|---|---|---|---|---|---|---|---|
| l | 28.16 | 22.32 | 18.23 | 21.6 | 16.67 | 18.7 | 16.2 | 20.91 | 21.46 |
| r | -0.35 | 8.48 | 9.4 | 8.51 | 7.44 | 9.61 | 7.46 | 9.41 | 5.09 |
| φ | 0 | 0 | 163 | 170.4 | 237 | 270 | 310 | 353 | 41.25 |

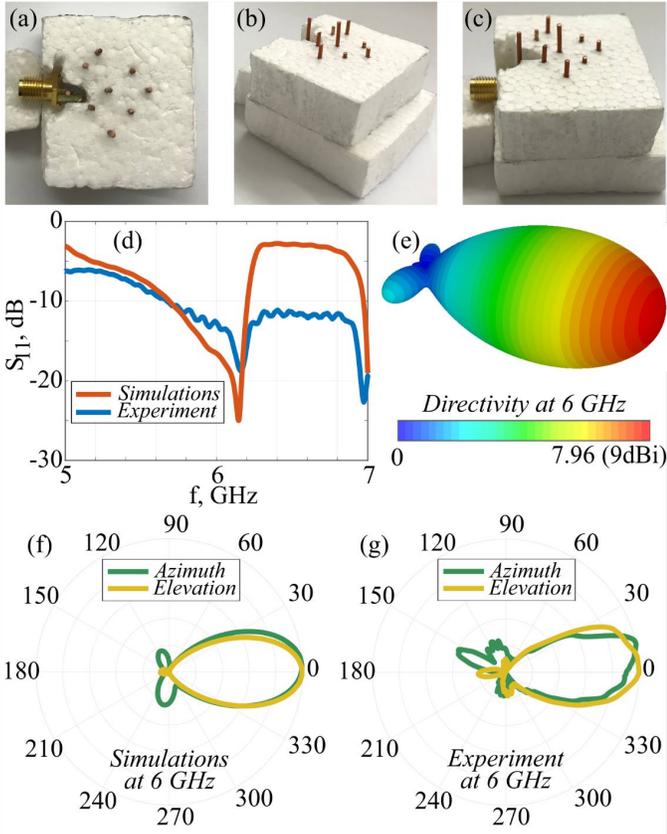

Fig. 2. (a-c) Photographs of the antenna from different perspectives. (d) S11-parameter spectra - numerical (red) and experimental (blue) curves. (e) 3D radiation pattern, numerical result. (e, f) radiation patterns in azimuthal (green) and elevation (yellow) cuts – numerical and experimental results. Patterns are plotted for 6 GHz, where the antenna directivity is the highest.

To characterize the results quantitatively, the following definitions are in use [1]:

$$D_\varphi = \frac{P_{max}}{\frac{1}{2\pi}\int_0^{2\pi} P(\varphi)d\varphi} \quad (1.1)$$

$$D_\theta = \frac{P_{max}}{\frac{1}{2\pi}\int_0^{2\pi} P(\theta)d\theta} \quad (1.2)$$

$$D = \frac{P_{max}(\varphi)+P_{max}(\theta)}{\frac{1}{4\pi}\left[\int_0^{2\pi} P(\varphi)d\varphi+\int_0^{2\pi} P(\theta)d\theta\right]} \quad (1.3)$$

where $D_\varphi$ and $D_\theta$ stay for directives in azimuthal and elevation cut planes, while $D$ is the total directivity of the antenna. $P$ is the far-field radiated power density. Table 2 summarizes the results. For an assessment, 9 dBi directivity can be achieved with an electrically large Yagi-Uda antenna, encompassing 6-7 elements.

Numerical predictions underestimate experimental values, which is quite atypically. The contributing element here is the feeding cable. In the numerical assessment, a discrete port was used to excite the system, neglecting the cable impact. However, the current flow on the cable does not introduce a significant change, and, consequently, implementing a balun is not mandatory. Apart from the directivity, the realized gain has been measured and the result appears in Table 2. To calibrate this measurement, IDPH-2018 S/N-0807202 horn antenna has been used. In the calibrated experiment, the measured gain is below the estimated numerical prediction.

TABLE 2.
Directivities and Gain – numerical and experimental results at 6 GHz.

|  | Numerical Model | Experiment |
|---|---|---|
| $D_\varphi$, dBi | 5.95 | 6.44 |
| $D_\theta$, dBi | 7.39 | 5.54 |
| $D$, dBi | 9.01 | 9.15 |
| $G$, dBi | 8.81 | 6.15 |

To further explore the radiation characteristics of the antenna, its directivity and gain spectra are plotted (Fig. 3). It is quite remarkable that the antenna maintains its radiation properties over an extended frequency band. It is a result of two main contributing factors. First, the antenna, nevertheless it is electrically small, it is not deeply subwavelength. The second effect comes from the resonance cascading, where multiple resonances of the system do have a spectral overlap but cover a sufficient bandwidth owing their side-by-side mutual positioning. This effect will be analyzed next.

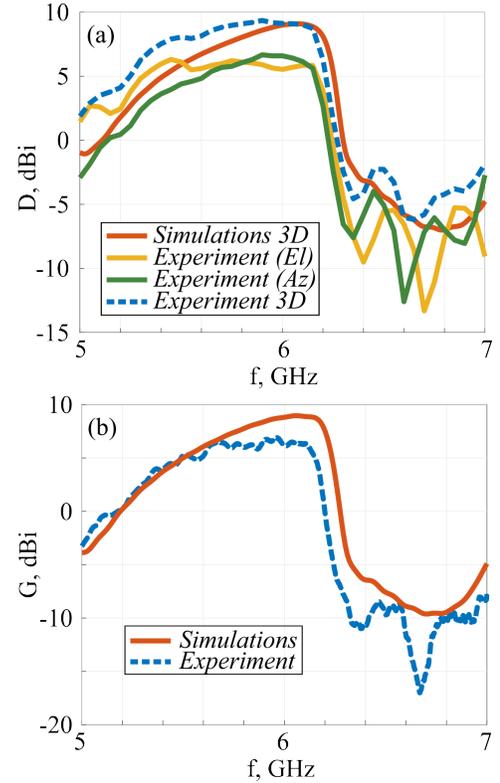

Fig. 3. (a) Azimuthal, elevation and total directives spectra – numerical and experimental results. (b) Gain spectrum – numerical (red) and experimental (blue).

IV. MULTIPOLE EXPANSION

Superdirectivity properties come from constructive interference of several resonant multipoles, which sum up into a directive radiation pattern. To demonstrate the effect, the multipole expansion of the radiation pattern will be performed next. There are a few possible approaches for analyzing electromagnetic scattering and radiation problems in various systems, e.g. [40]–[51]. Fig. 4 demonstrates the spectra of several lower-order multipoles. The convergence of the multipolar sum to the total radiated power is assessed. First, it can be seen that the six lowest multipoles are sufficient to describe the antenna radiation. Second, the multipoles resonate at the vicinity of 6GHz and cover a 13% fractional bandwidth, as the resonances are placed side by side. This resonant cascading approach to bandwidth extension allows superdirective antennas to support wireless links within existing communication standards and motivates applying the proposed methodology to miniaturization.

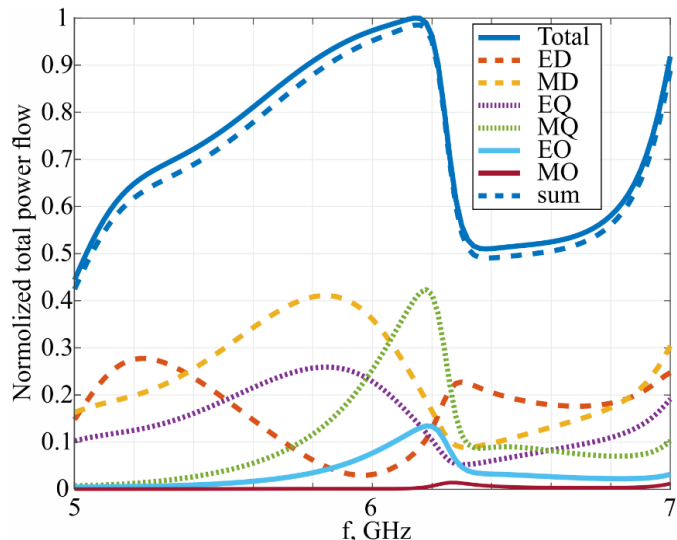

Fig. 4. Multipole expansion of the antenna radiation pattern. ED, MD – electrical and magnetic dipoles, EQ, MQ – quadrupoles and EO, MO – octupoles contributions.

## V. Performance assessment

After revealing the antenna performances and underlining the principles of its basic operation, the device can be assessed versus existing and widely used bounds and compared with other superdirective demonstrations. The assessment will be made versus Chu-Harrington and Geyi limits. Fig. 5 is the directivity versus an electrical size of device size ($2R/\lambda$, where R is the radius of an enclosing sphere and $\lambda$ is the operational wavelength). Several recently reported superdirective antennas were chosen for making an assessment [52]–[55]. Our device is well positioned above the Chu-Harrington and Geyi limits. The dots on the chart correspond to the values, reported in the literature. It is worth noting that in a vast majority of cases, bandwidth characteristics are not discussed as it is widely believed that a superdirective antenna has to be narrowband (we question this claim by demonstrating our architecture). To address the gain properties, we explored several reported designs and calculated operational bandwidth. Specifically, gain spectra were calculated and plotted versus $2R/\lambda$, the x-axis parameter in Fig. 5. Solid line in Fig. 5 correspond to the calculation. In the case of our design, directivity spectra (both numerical and experimental) are presented. According to the figure of merit, related to the bandwidth, our antenna has the best characteristics.

To summarize the bandwidth performances quantitatively, the superdirective bandwidth was defined versus both Chu-Harrington and Geyi limits – the frequency for which the directivity is above those bounds. Table 3 summarizes the results. Presents are calculated versus the central frequency in the range within the superdirective bandwidth. According to this definition, our realization has a significant advantage and, furthermore, makes it a legitimate hardware to support wireless communication links.

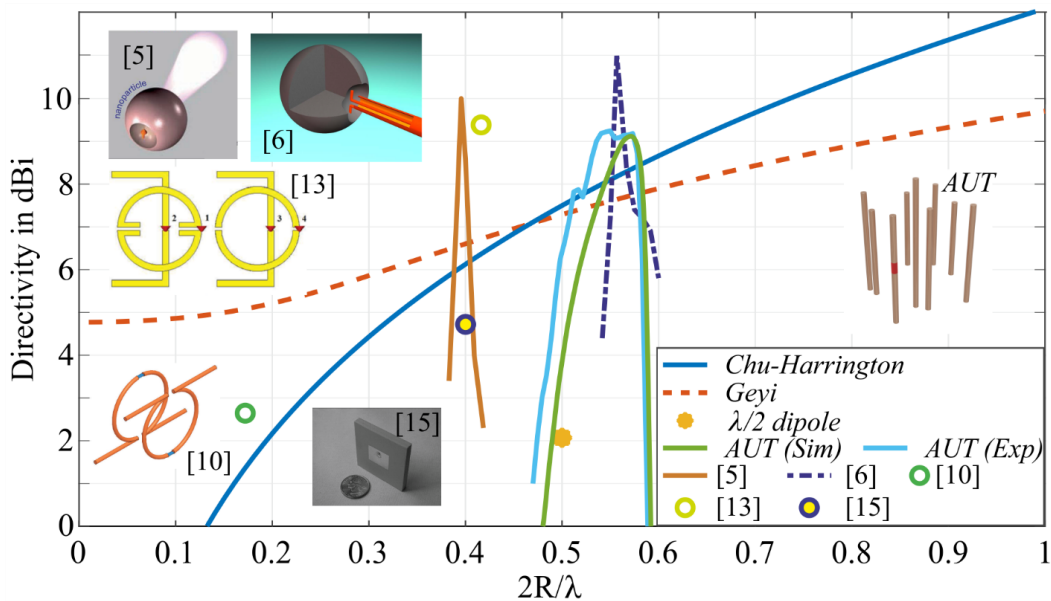

Fig. 5. Chu-Harrington and Geyi limits [52] versus an electrical size of an antenna. Points on the chart correspond to several reported realizations, where performances at a single frequency were reported (references are in the legend). Solid lines – directivity spectra, calculated from data, adopted from indicated references. Antenna under test (AUT) – our realization with both numerical and experimental results presented. Yellow dot - conventional $\lambda/2$ dipole antenna for reference.

TABLE 3.
BW of the superdirective broadband antennas over the fundamental limits (Fig. 5)

| Ref. | BW over Chu-Har., % | BW over Geyi, % |
|---|---|---|
| AUT (Sim) | 6.6 | 8.3 |
| AUT (Exp) | 12.2 | 13 |
| [5] | 4.4 | 3.8 |
| [6] | 2.7 | 3.7 |

VI. CONCLUSION

A supergain broadband antenna has been demonstrated. The concept of resonant cascading has been developed and employed towards achieving both high directivity and bandwidth operation within a subwavelength device. While a spectral collocation of a large number of resonant multipoles is responsible for obtaining highly directive patterns, tuning multipole hierarchy and spreading their resonant responses over a span of frequencies can grant both directivity and bandwidth. Here we demonstrate a compact electrically small antenna with 9dBi directivity and 13% fractional bandwidth at 6 GHz. Resonant cascading of six lower-order multipoles was shown to govern the radiation characteristics. The experimentally demonstrated gain is 6.15 dBi.

The capability to achieve high gain properties within a small footprint device and without a bandwidth degradation makes the proposed superdirective elements to be very attractive in many wireless applications, owing to new strategies in hardware elements miniaturization.

VII. ACKNOWLEDGMENTS